\documentstyle[psfig]{l-aa}

\def\AJ{AJ }
\def\ApJ{ApJ }
\def\AA{A\&A }
\def\AAS{A\&AS }
\def\ARAA{ARA\&A }

\def\ApJS{ApJS }

\def\MNRAS{MNRAS }
\def\amm{NH$_{3}$ }

\def\hc5n{HC$_5$N }

\begin{document}
\thesaurus{08(09.03.01; 09.13.2; 13.19.3; 09.09.1 TMC)}
\title{Four new dense molecular cores in the Taurus Molecular Cloud (TMC)}   
\subtitle{Ammonia and cyanodiacetylene observations}
\author{C. Codella\inst{1}, R. Welser\inst{1}, C. Henkel\inst{1},
P.J. Benson\inst{2} \and P.C. Myers\inst{3}}
\offprints{C. Henkel}
\institute{Max-Planck-Institut f\"ur Radioastronomie, Auf dem H\"ugel 69,
53121 Bonn, Germany
\and
Whitin Observatory Wellesley, Wellesley, Massachusetts 02181
\and
Harvard-Smithsonian Center for Astrophysics, 60 Garden Street, Cambridge,
MA 02138}
\date{Received date; accepted date}
\maketitle
\markboth{Codella et al.: Four new dense molecular cores in the TMC}{ }
\begin{abstract}

Trying to obtain a more complete picture of star forming regions in the Taurus
molecular cloud, four newly discovered dense molecular cores (L1521D, L1521F,
L1524, L1507A) are identified and mapped through the ammonia (J,K) = (1,1) and
(2,2) rotational inversion lines. These cores have sizes from 0.06 to 0.09 pc,
hydrogen densities from 0.6~10$^{4}$ to 19.9~10$^{4}$ cm$^{-3}$ and kinetic
temperatures between 7.9 and 9.9 K. The masses range from 0.2 and 1.0 $M_{\rm
\sun}$, placing the cores with the lowest $M$ values at the lower edge of the
mass distribution for ammonia cores in the TMC. Turbulent, thermal and
gravitational energies have been estimated. A comparison between these energy
terms and considerations related to thermal and turbulent line broadening
indicate that these cores are close to gravitational equilibrium. Moreover, we
report detections of the J=9--8 transition of \hc5n towards the ammonia peak
positions of these four molecular cores. The HC$_5$N column densities are
between 1.6 10$^{12}$ and 9.2 10$^{12}$ cm$^{-2}$, in agreement with values 
derived for other molecular cores located in the TMC.

\end{abstract}
\section{Introduction}

The Taurus molecular cloud (TMC) is one of the best targets to study low-mass
star formation, since it is located at a distance of only 140 pc. The region
is associated with T Tauri stars, identified by infrared and optical
observations (e.g. Strom et al. \cite{strom}; Kenyon et al. \cite{kenyon};
Weintraub \cite{wein}), low-luminosity objects extracted from the IRAS Point
Source Catalogue (PSC, IRAS \cite{iras}), molecular outflows (e.g. Fukui
\cite{fukui1}; Fukui et al. \cite{fukui2}) and water maser sources (Wouterloot
et al. \cite{wbf}), indicating ongoing star formation. The TMC contains
several dark molecular clouds of size $\simeq$ 1 pc and densities of 10$^{3}$
cm$^{-3}$ (e.g. Walmsley et al. \cite{wal}; Wouterloot \& Habing
\cite{wouhab}; Ungerechts \& Thaddeus \cite{unghetad}; Mizuno et al.
\cite{mizuno}). These in turn contain dense molecular cores, high density
($\ge$ 10$^{4}$ cm$^{-3}$) substructures which have been classified by Myers
et al. (\cite{mylibe}) and Myers \& Benson (\cite{mybe}). For the sky
distribution of the molecular cores in the TMC, see Kenyon et al.
(\cite{kenyon}). Subsequent multimolecular observations (e.g. Benson \& Myers
\cite{bemy}; Fuller \cite{fuller}; Fuller \& Myers \cite{fumy}) have confirmed
that dense cores are smaller than 0.1 pc with masses ranging from 0.5 to 10
solar masses and kinetic temperatures of the order of 10 K. Beichman et al.
(\cite{beich1}; \cite{beich2}) have shown that about 50\% of the molecular
cores are associated with IRAS sources, indicating that these objects may
contain, in their interiors, young stellar objects (YSOs). Therefore, it is
now clear that there exists a close association between low-mass YSOs and
molecular clumps. Since the TMC is not strongly affected by the nearby Cas-Tau
OB association (Blaauw \cite{blaauw}), it is clear that it provides unique
opportunities to study spontaneous low-mass star formation and, in particular,
to investigate the nature of the earliest stages of low-mass star formation,
strictly connected with dense molecular clumps.

Following the definition given by Myers \& Benson (\cite{mybe}; i.e. sources
with $T^{*}_{A}$ $\ge$ 0.4 K at the Haystack and Green Bank antennas), 16
dense molecular cores have so far been identified in the TMC (Myers \& Benson
\cite{mybe}; Benson \& Myers \cite{bemy2}). In order to obtain a more complete
sample of TMC cores, to gain information about the nature of the molecular gas
before or at the onset of YSO formation and to clarify the evolutionary stages
connected with the collapse process, we report observations of the \amm
($J,K$) = (1,1) and (2,2) inversion lines and the cyanodiacetylene (HC$_5$N)
J=9--8 rotational line towards 4 newly discovered dense cores.

\section{Selection criteria}

The molecular cores chosen for this project have been selected from an
extensive search for ammonia emission towards 26 high visual extinction
regions in the TMC identified from the Palomar Observatory Sky Survey (POSS)
maps and from Myers et al. (\cite{myhobe}). The observed positions are located
within 04$^{h}$ 15$^{m}$ $\le$ $\alpha$ $\le$ 05$^{h}$ 00$^{m}$ and +16$\degr$
$\le$ $\delta$ $\le$ +30$\degr$: 9 regions in the long filament which includes
TMC-2 (Kutner's Cloud), 7 regions in the parallel filament to the north,
containing B217 (see Fig. 1 of Walmsley et al. \cite{wal}), and 10
isolated optically obscured regions where no extended clouds were seen on the
POSS plates. For the survey, the Haystack (USA) 37-m and the Green Bank (USA)
NRAO 43-m antennas have been used with FWHP (full width half power) beam sizes
of 84$\arcsec$ and beam efficiencies of about 0.25 during several runs in
1983.

Table 1 summarizes the regions where \amm line emission was searched for. For
the non-detections the r.m.s. $T^{*}_{A}$ value is reported. Ammonia emission
has been detected towards 9 sources. Out of these 9, 5 can be classified as
dense molecular cores: L1521D, L1521F, L1524, L1535 and L1507A. Since L1535
has already been investigated by Ungerechts et al. (\cite{unghe}), we have
focused our attention on the other four cores, which have been mapped in
NH$_3$, using the 100-m MPIfR radio-telescope at Effelsberg (Germany).

\begin{table}
\caption[] {Regions investigated through NH$_3$ observations}
\begin{tabular}{r|r|r|c|l}
\hline 
\multicolumn{1}{c|}{\bf {$\alpha$}} &
\multicolumn{1}{c|}{\bf {$\delta$}} & 
\multicolumn{1}{c|}{\bf {$T_{\rm A}^{\rm *}$}} &
\multicolumn{1}{c|}{{Antenna $^{(a)}$}} &
\multicolumn{1}{c}{Dense core} \\ 
\multicolumn{1}{c|}{(1950)} &
\multicolumn{1}{c|}{(1950)} &
\multicolumn{1}{c|}{(mK)} &
\multicolumn{1}{c|}{ } &
\multicolumn{1}{c}{ } \\
\hline
04 17 56 & +26 55 23 &     449 & GB & L\,1521\,D \\
04 18 55 & +29 21 00 & $<$ \hspace*{0.16cm}95 & HA \\
04 19 55 & +19 19 49 & $<$ 102 & GB \\
04 20 30 & +26 32 47 & $<$ 101 & GB \\
04 21 48 & +27 10 58 & $<$ \hspace*{0.16cm}95 & GB \\ 
04 23 36 & +24 32 40 & $<$ 150 & HA \\
04 25 00 & +18 35 12 & $<$ 103 & GB \\
04 25 35 & +26 45 00 &     977 & GB & L\,1521\,F \\
04 26 20 & +24 28 06 &     603 & GB & L\,1524 \\
04 26 25 & +26 52 47 & $<$ 102 & GB \\
04 27 58 & +24 21 06 & $<$ 113 & HA \\
04 28 02 & +24 05 20 & $<$ 108 & HA \\
04 28 40 & +27 02 13 & $<$ \hspace*{0.16cm}85 & GB \\
04 29 38 & +24 42 13 & $<$ \hspace*{0.16cm}80 & HA \\
04 29 40 & +24 06 00 & $<$ 151 & HA \\
04 29 57 & +24 10 53 & $<$ 179 & HA \\
04 31 08 & +24 02 10 &     216 & GB \\
04 32 37 & +24 01 42 &     788 & HA & L\,1535 \\
04 37 28 & +29 50 00 & $<$ \hspace*{0.16cm}74 & HA \\
04 37 39 & +29 48 37 &     270 & GB \\
04 39 29 & +29 38 07 &     472 & GB & L\,1507\,A \\
04 40 20 & +29 34 37 &     276 & GB \\
04 42 18 & +16 31 58 & $<$ \hspace*{0.16cm}85 & GB \\
04 42 22 & +17 01 42 & $<$ \hspace*{0.16cm}86 & GB \\
04 42 38 & +16 51 35 & $<$ \hspace*{0.16cm}85 & GB \\
04 57 16 & +26 10 23 &     236 & GB \\
\hline
\end{tabular}
\vskip 0.5cm
$^{(a)}$ Observed with the 43-m Green Bank antenna (GB) or
with the 37-m Haystack antenna (HA) in 1983. The FWHP beam sizes are 
$\simeq$ 84 arcsec, while the beam efficiencies are $\simeq$ 0.25. 
\end{table}

\section{Observations}

The observations were made during several runs in 1985, 1986 and 1996 with the
MPIfR antenna employing a maser receiver during the 1985-1986 and a HEMT
receiver during the 1996 periods. At the rest frequency of the ($J,K$) = (1,1)
and (2,2) lines of \amm (23694.496 MHz and 23722.631 MHz, respectively) and of
the J=9--8 line of HC$_5$N (23963.888 MHz) the FWHP beam size is 40$\arcsec$.
The zenith system temperature was around 60 K in good weather conditions on an
atmosphere corrected antenna temperature ($T_{\rm A}^{\rm *}$) scale. The
observations were made using position switching, with an integration time of
typically 3 minutes on source. Spectral information was obtained with a
1024--channel three--level autocorrelation spectrometer. We observed the \amm
(1,1) and (2,2) lines simultaneously by splitting the autocorrelator into two
spectrometers of 512 channels with bandwidths of 6.25 MHz and channel spacings
of 0.154 km s$^{-1}$. The velocity resolution was 0.185 km s$^{-1}$, i.e. a
factor 1.2 larger than the channel spacing. We observed the HC$_5$N line with
the autocorrelator split into two parts, each with a bandwidth of 3.12 MHz and
a velocity resolution of 0.092 km s$^{-1}$. The spectra from the two split
parts were then co-added in order to reduce the noise. The intensity scale of
the spectra was calibrated on the continuum sources NGC7027, 3C123 and 3C286,
according to the flux densities given by Baars et al. (\cite{baars}). The
pointing accuracy was better than 10$\arcsec$, corresponding to a point source
flux density error of $<$ 15\%. The calibration accounted for the dependence
of the gain on the elevation. Total flux density uncertainties are of the
order of 20\%.

\section{Results}

\begin{figure}
\centerline{\psfig{figure=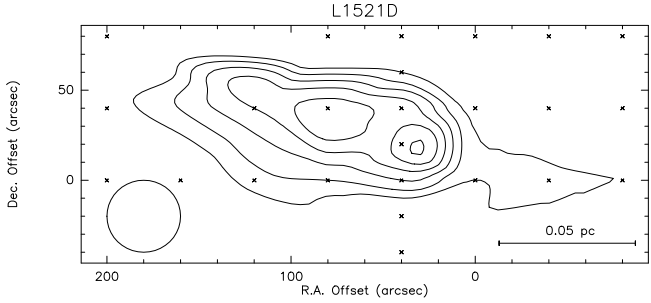,width=9.2cm,angle=-90}}
\caption[]{Contour \amm (1,1) $T_{\rm MB}$ map of L1521D. The (0,0) position
corresponds to: $\alpha$ = 04$^{h}$ 17$^{m}$ 56$^{s}$; $\delta$ = +26$\degr$
55$\arcmin$ 23$\arcsec$. The contour levels are from 1.4 K to 2.4 K by steps
of 0.2 K ($\simeq$ 1 r.m.s. noise). The empty circle shows the Effelsberg beam
(HPBW), the small crosses mark measured positions.}
\end{figure}

\begin{figure}
\centerline{\psfig{figure=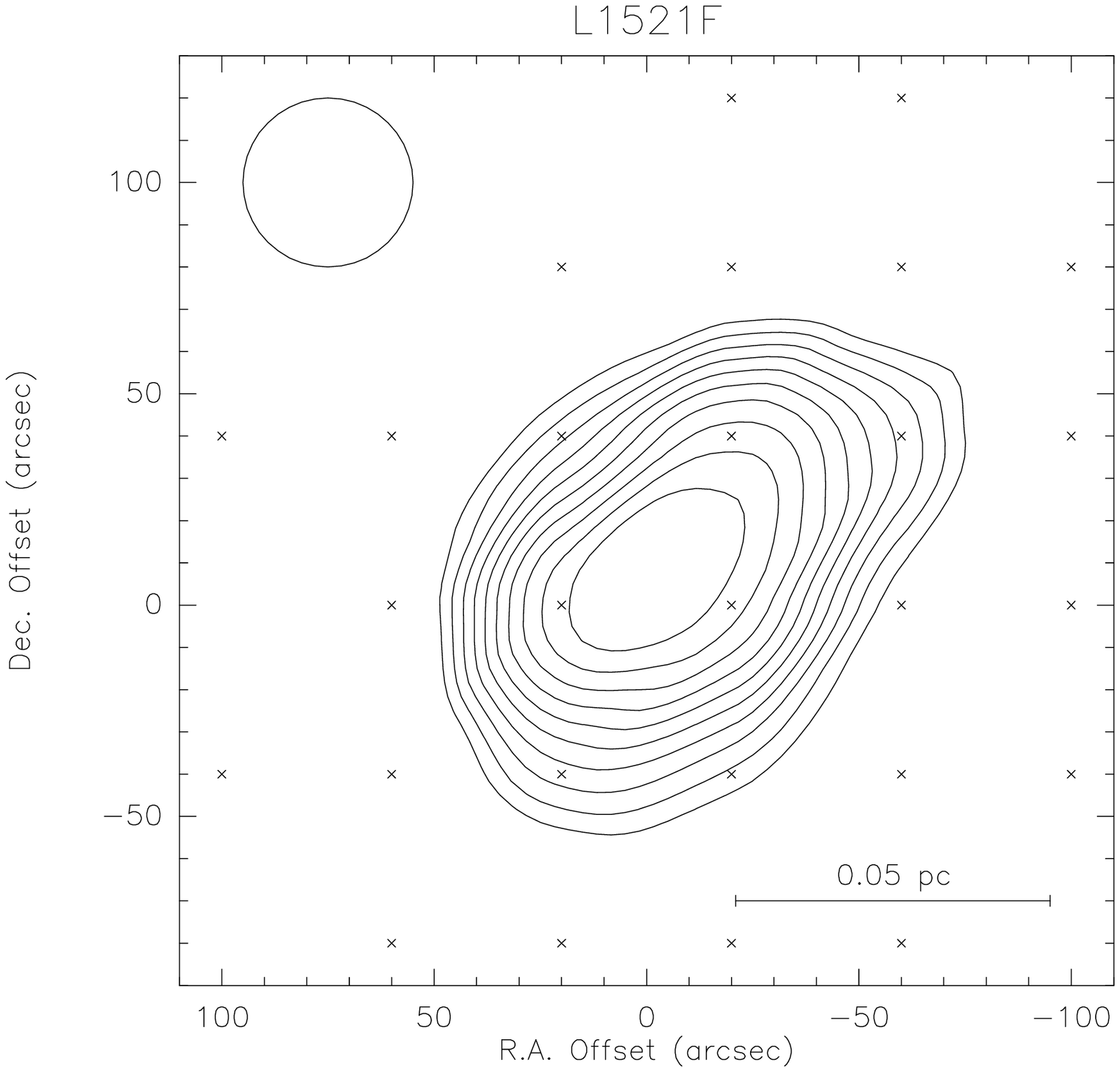,width=9.2cm,angle=-90}}
\caption[]{Contour \amm (1,1) $T_{\rm MB}$ map of L1521F. The (0,0) position
corresponds to: $\alpha$ = 04$^{h}$ 25$^{m}$ 35$^{s}$; $\delta$ = +26$\degr$
45$\arcmin$ 00$\arcsec$. The contour levels are from 2.6 K to 6.4 K by steps
of 0.4 K ($\simeq$ 2 r.m.s. noise). The empty circle shows the Effelsberg beam
(HPBW).}
\end{figure} 

\begin{table*}
\caption[] {Observed \amm (1,1) line parameters. The (1950) coordinates of the
peak positions of the four cores are: ($\alpha$ = 04$^{h}$ 17$^{m}$ 59$^{s}$;
$\delta$ = +26$\degr$ 55$\arcmin$ 43$\arcsec$) and ($\alpha$ = 04$^{h}$
18$^{m}$ 02$^{s}$; $\delta$ = +26$\degr$ 56$\arcmin$ 03$\arcsec$) for L1521D,
($\alpha$ = 04$^{h}$ 25$^{m}$ 36$^{s}$; $\delta$ = +26$\degr$ 45$\arcmin$
00$\arcsec$) for L1521F, ($\alpha$ = 04$^{h}$ 26$^{m}$ 21$^{s}$; $\delta$ =
+24$\degr$ 28$\arcmin$ 26$\arcsec$) for L1524 and ($\alpha$ = 04$^{h}$
39$^{m}$ 29$^{s}$; $\delta$ = +29$\degr$ 38$\arcmin$ 17$\arcsec$) for L1507A}
\begin{tabular}{l|rr|r|r|r|r|r}
\hline
\multicolumn{1}{c|}{{Name}} &
\multicolumn{1}{c}{\bf {$\Delta$$\alpha$}} &
\multicolumn{1}{c|}{\bf {$\Delta$$\delta$}} &
\multicolumn{1}{c|}{\bf {$T_{\rm MB}$}} &
\multicolumn{1}{c|}{\bf {$v_{\rm LSR}$}} &
\multicolumn{1}{c|} {{FWHM}} &
\multicolumn{1}{c|} {\bf {$\tau_{\rm tot}$}} &
\multicolumn{1}{c}{\bf {$\Delta v_{\rm INT}$}} \\
\multicolumn{1}{c|}{} &
\multicolumn{1}{c}{(~${''}$~)} &
\multicolumn{1}{c|}{(~${''}$~)} &
\multicolumn{1}{c|}{(K)} &
\multicolumn{1}{c|}{(km s$^{-1}$)} &
\multicolumn{1}{c|}{(km s$^{-1}$)} &
\multicolumn{1}{c|}{} &
\multicolumn{1}{c}{(km s$^{-1}$)} \\
\hline
L1521D &  +40,& +20 & 2.3 (0.1) & 6.94 (0.15) & 0.8 (0.2) &  4.5 (1.1) & 0.33 (0.02) \\
       &  +80,& +40 & 2.3 (0.1) & 6.97 (0.15) & 0.6 (0.2) &  4.6 (1.3) & 0.30 (0.02) \\
L1521F &  +20,&   0 & 6.1 (0.3) & 6.62 (0.15) & 0.7 (0.2) & 12.0 (0.8) & 0.25 (0.01) \\
L1524  &  +20,& +20 & 4.7 (0.2) & 6.40 (0.15) & 0.8 (0.2) & 14.0 (0.9) & 0.26 (0.01) \\
L1507A &    0,& +10 & 3.4 (0.2) & 6.16 (0.15) & 0.7 (0.2) &  5.8 (0.9) & 0.27 (0.01) \\
\hline
\end{tabular}
\end{table*}

Ammonia maps have been obtained of the four dense cores L1521D, L1521F, L1524
and L1507A in the ($J,K$) = (1,1) line (Figs. 1 to 4). In order to obtain the
observed source size, we have determined the half maximum contours of the
ammonia main beam brightness temperature ($T_{\rm MB}$; K) and have calculated
the apparent geometric mean diameter (FWHP) $D$ = $\sqrt{ab}$, where $a$ and
$b$ are the major and minor-axis, respectively. The ammonia core sizes are
0.09 pc (L1521D), 0.07 pc (L1521F), 0.08 pc (L1524) and 0.06 pc (L1507A). 
Deconvolution by the 0.027 pc beam does not significantly alter these values.
For those spectra where the main and satellite hyperfine components have been
observed with a sufficient signal-to-noise ratio ($>$ 3) we have obtained the
total optical depth ($\tau_{tot}$) and the intrinsic linewidth (FWHM) of the
hyperfine components ($\Delta$v$_{\rm INT}$; km s$^{-1}$) using a program
which fits these parameters to the observed spectra, assuming gaussian
profiles in an individual component and relative opacities being consistent with
the Local Thermodynamic Equilibrium (LTE) approximation. For the positions
where the satellite components have intensities below the noise level, but the
main component of the \amm (1,1) line has been detected, we have fitted a
single gaussian profile. Moreover, we have observed the \amm (2,2) transition
towards the peak positions of the four molecular cores.

\begin{table}
\caption[] {Observed \amm (2,2) line parameters}
\begin{tabular}{l|rr|r|r|r}
\hline
\multicolumn{1}{c|}{{Name}} &
\multicolumn{1}{c}{\bf {$\Delta$$\alpha$}} &
\multicolumn{1}{c|}{\bf {$\Delta$$\delta$}} &
\multicolumn{1}{c|}{\bf {$T_{\rm MB}$}} &
\multicolumn{1}{c|}{\bf {$v_{\rm LSR}$}} &
\multicolumn{1}{c} {{FWHM}} \\
\multicolumn{1}{c|}{} &
\multicolumn{1}{c}{(~${''}$~)} &
\multicolumn{1}{c|}{(~${''}$~)} &
\multicolumn{1}{c|}{(K)} &
\multicolumn{1}{c|}{(km s$^{-1}$)} &
\multicolumn{1}{c}{(km s$^{-1}$)} \\
\hline
L1521D &  +40,& +20 & 0.5 (0.1)  & 6.81 (0.15) & 0.6 (0.2) \\
L1521F &  +20,&   0 & 1.3 (0.2)  & 6.65 (0.15) & 0.3 (0.2) \\
L1524  &  +20,& +20 & 0.9 (0.2)  & 6.44 (0.15) & 0.3 (0.2) \\
L1507A &    0,& +10 & 1.1 (0.3)  & 6.17 (0.15) & 0.1 (0.2) \\
\hline
\end{tabular}
\end{table}

Table 2 displays the observed \amm (1,1) line parameters regarding the peak
positions: the source name, the right ascension and declination offset
($\arcsec$) of the peak position with respect to the coordinates given in
Table 1, the main beam brightness temperature, the LSR velocity ($v_{\rm
LSR}$; km s$^{-1}$), the apparent line width (FWHM; km s$^{-1}$) of the main
component, the total optical depth and the intrinsic linewidth of the
hyperfine components including instrumental broadening. The errors (r.m.s.) of
the derived parameters are given in parentheses. Figure 5 shows the \amm (1,1)
and (2,2) spectra of the peak position of each mapped source. Table 3 lists
the observed \amm (2,2) gaussian line parameters towards the \amm (1,1) peak
positions: the main beam brightness temperature, the LSR velocity and the FWHM
line width including instrumental broadening. For L1521D, we report the
observed parameters towards the (+40, +20) peak position, since the integrated
ammonia emission associated with this position is slightly larger that that of
the (+80, +40) peak position.

\begin{figure}
\centerline{\psfig{figure=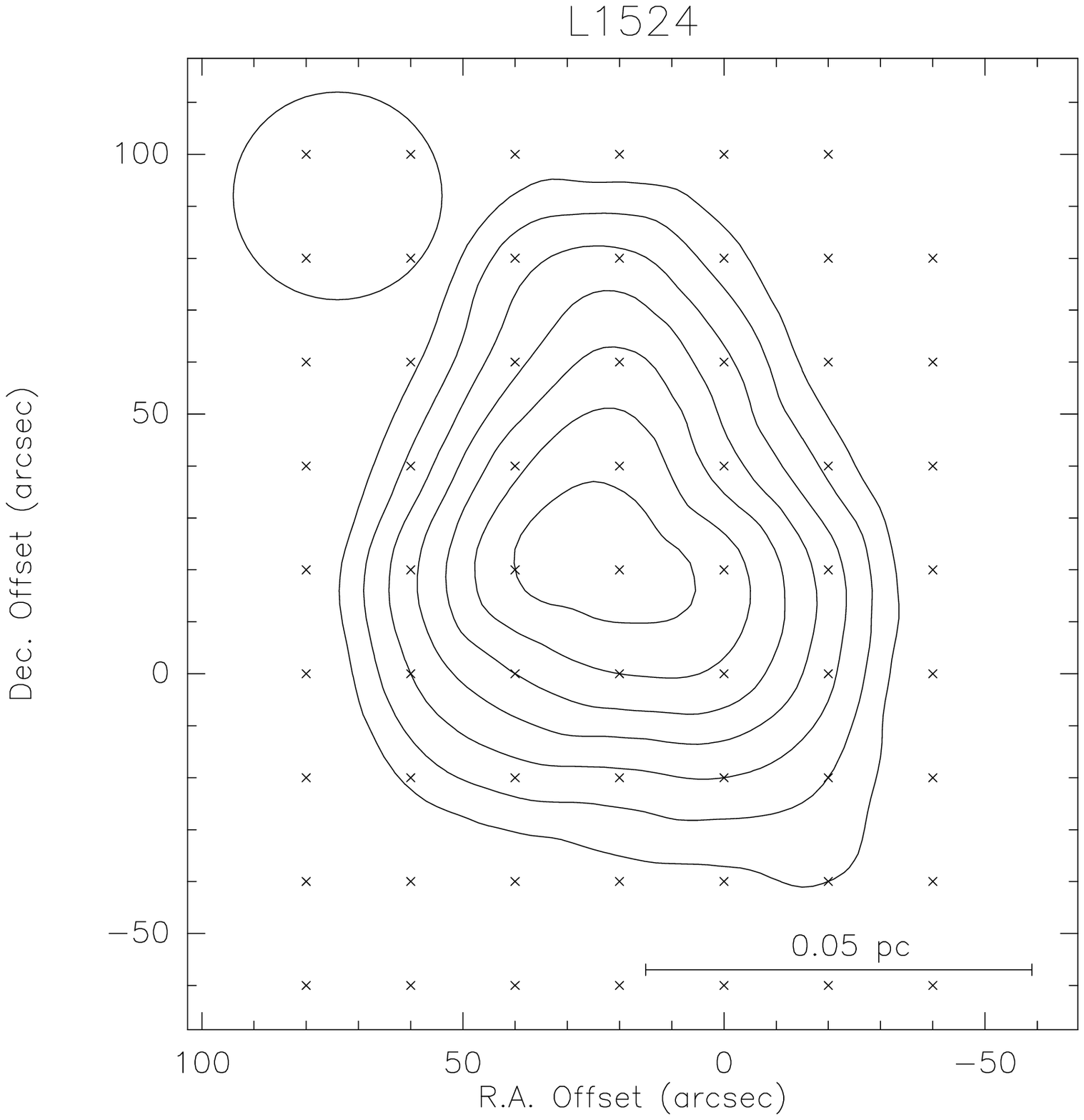,width=9.2cm,angle=-90}}
\caption[]{Contour \amm (1,1) $T_{\rm MB}$ map of L1524. The (0,0) position
corresponds to: $\alpha$ = 04$^{h}$ 26$^{m}$ 20$^{s}$; $\delta$ = +24$\degr$
28$\arcmin$ 06$\arcsec$. The contour levels are from 2.1 K to 4.5 K by steps
of 0.4 K ($\simeq$ 2 r.m.s. noise). The empty circle shows the Effelsberg beam
(HPBW).}
\end{figure}

\begin{figure}
\centerline{\psfig{figure=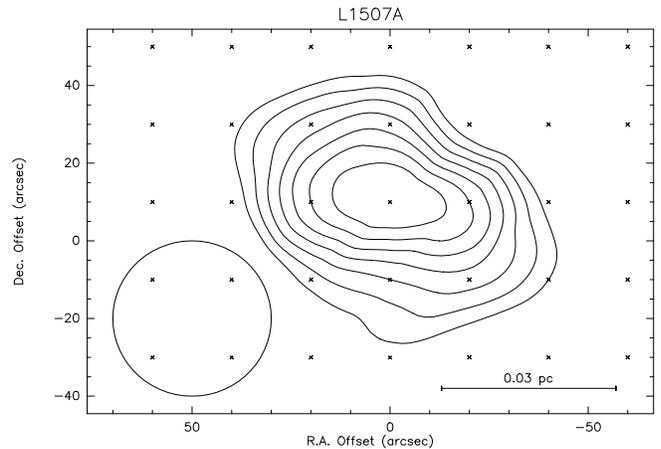,width=9.2cm,angle=-90}}
\caption[]{Contour \amm (1,1) $T_{\rm MB}$ map of L1507A. The (0,0) position
corresponds to: $\alpha$ = 04$^{h}$ 39$^{m}$ 29$^{s}$; $\delta$ = +29$\degr$
38$\arcmin$ 07$\arcsec$. The contour levels are from 2.0 K to 3.4 K by steps
of 0.2 K ($\simeq$ 1 r.m.s. noise). The empty circle shows the Effelsberg beam
(HPBW).}
\end{figure}

Table 4 gives the HC$_5$N (J=9--8) line parameters towards the ammonia peak
positions, obtained from gaussian fits: the parameters are main beam
brightness temperature, LSR velocity and FWHM line width including
instrumental broadening. Figure 6 shows the HC$_5$N (J=9--8) spectra.

\begin{figure*}
\centerline{\psfig{figure=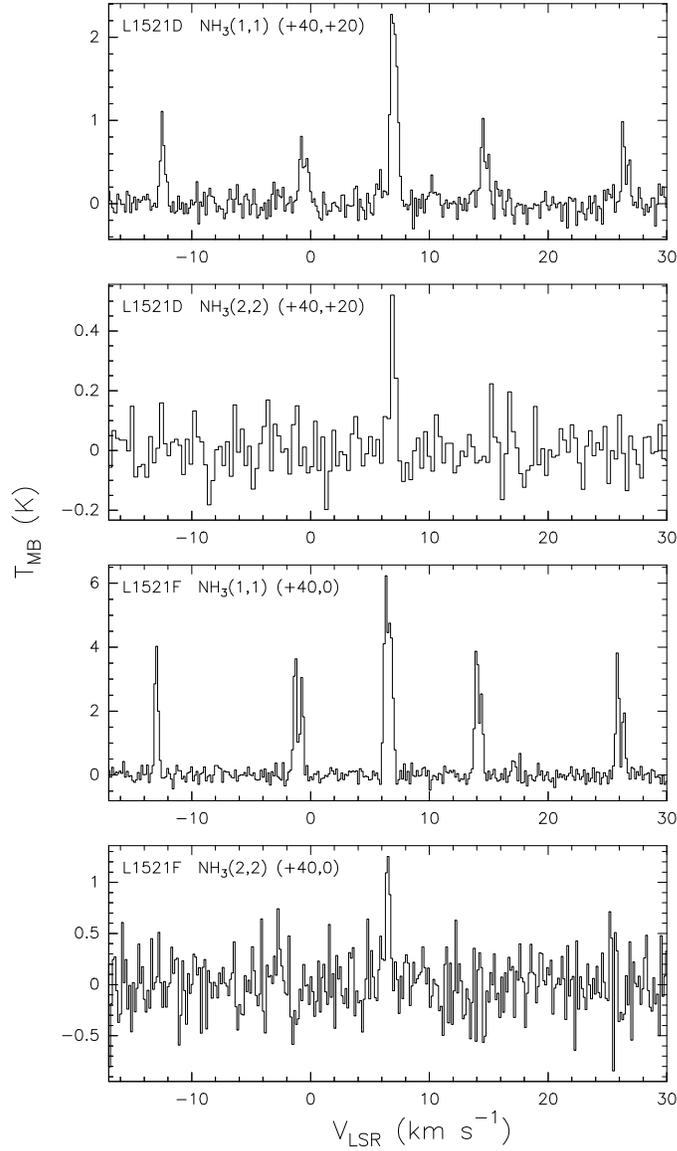,width=10cm}}
\caption[]{Ammonia (1,1) and (2,2) spectra of the peak position of each mapped
core. Source name and angular offset map position (in arcseconds) are
indicated.}
\end{figure*}

\section{Discussion}

\begin{figure*}
\centerline{\psfig{figure=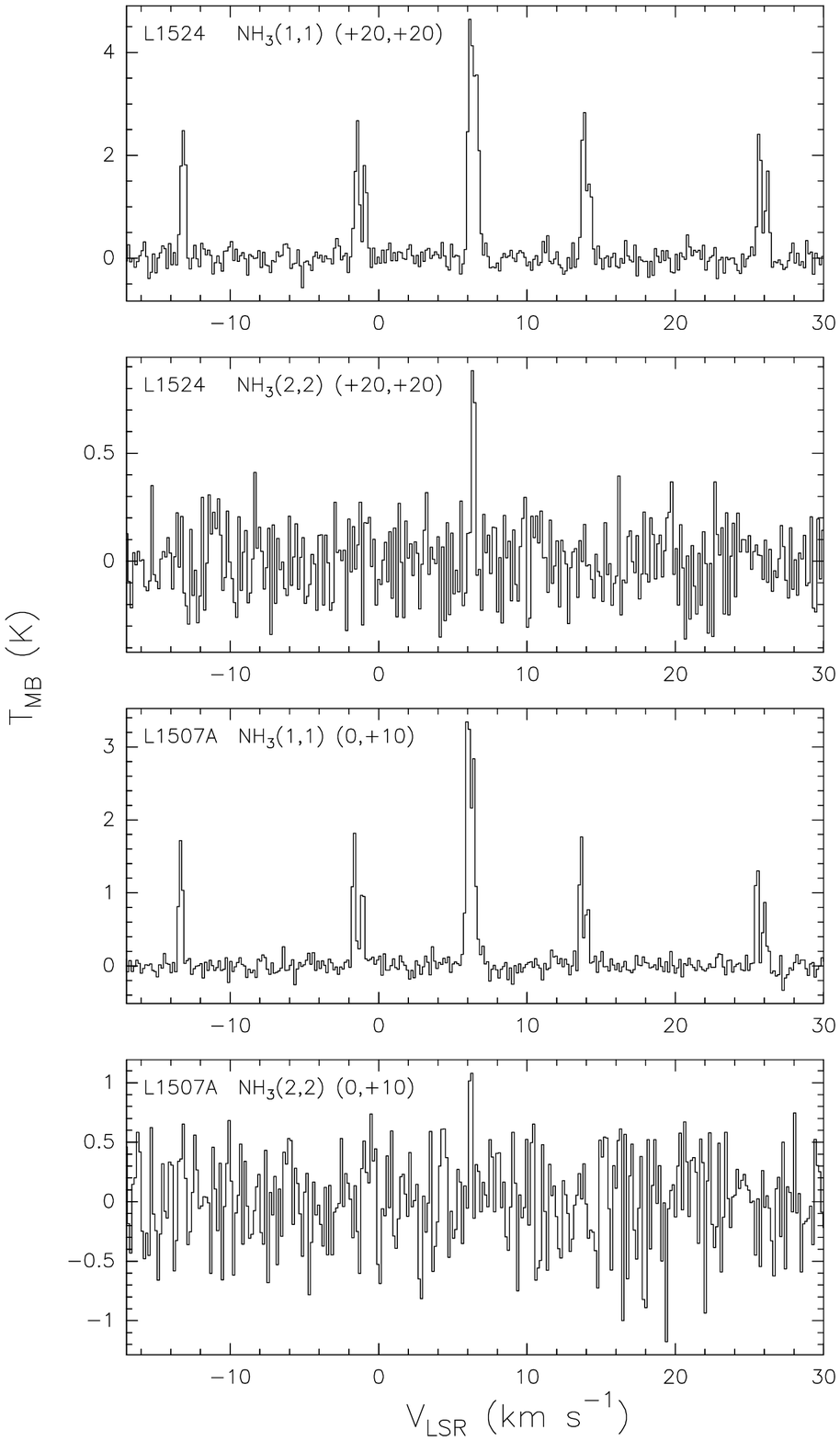,width=10cm}}
\addtocounter{figure}{-1}
\caption[]{(continued)}
\end{figure*}

\begin{figure*}
\centerline{\psfig{figure=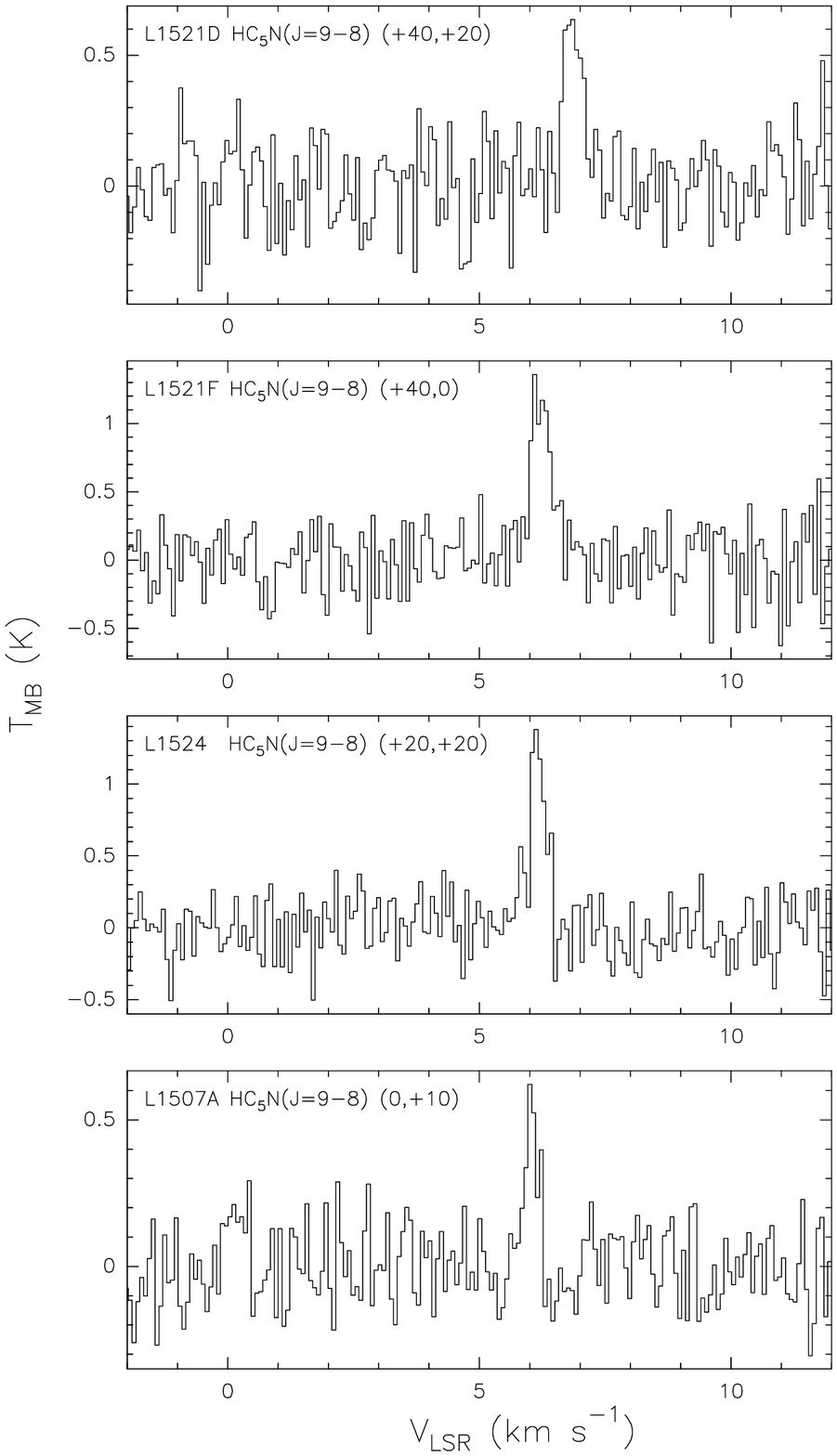,width=10cm}}
\caption[]{Cyanodiacetylene (J=9--8) spectra at the ammonia peak position of 
each molecular core. Source name and angular offset (in arcseconds) 
relative to the (0,0) positions in the ammonia maps (see Figs. 1-4) are 
indicated.}
\end{figure*}

\begin{table*}
\caption[] {Observed and derived HC$_5$N (J=9--8) line parameters}
\begin{tabular}{l|rr|r|r|r|r}
\hline 
\multicolumn{1}{c|}{{Name}} &
\multicolumn{1}{c}{\bf {$\Delta$$\alpha$}} &
\multicolumn{1}{c|}{\bf {$\Delta$$\delta$}} &
\multicolumn{1}{c|}{\bf {$T_{\rm MB}$}} &
\multicolumn{1}{c|}{\bf {$v_{\rm LSR}$}} & 
\multicolumn{1}{c|} {{FWHM}} &
\multicolumn{1}{c}{\bf {$N_{\rm HC_5N}$}} \\ 
\multicolumn{1}{c|}{} &
\multicolumn{1}{c}{(~${''}$~)} &
\multicolumn{1}{c|}{(~${''}$~)} &
\multicolumn{1}{c|}{(K)} &
\multicolumn{1}{c|}{(km s$^{-1}$)} &
\multicolumn{1}{c|}{(km s$^{-1}$)} &
\multicolumn{1}{c}{(10$^{12}$ cm$^{-2}$)} \\ 
\hline
L1521D &  +40,& +20 & 0.6 (0.2) & 6.86 (0.09) & 0.4 (0.1) & 4.5 (0.7) \\ 
L1521F &  +20,&   0 & 1.2 (0.2) & 6.22 (0.09) & 0.5 (0.1) & 9.2 (1.0) \\
L1524  &  +20,& +20 & 1.3 (0.2) & 6.16 (0.09) & 0.4 (0.1) & 7.5 (0.9) \\
L1507A &    0,& +10 & 0.6 (0.1) & 6.03 (0.09) & 0.3 (0.1) & 1.6 (0.3) \\
\hline
\end{tabular}
\end{table*}

\subsection{Derived parameters from \amm observations}

The rotational temperature ($T_{\rm rot}$) is defined by applying the
Boltzmann distribution to the populations of different (J,K) rotational levels
and has been calculated from the (1,1) and (2,2) data following the procedure
outlined by H\"uttemeister et al. (\cite{hutte}) and Lemme (\cite{lemme2}).
Given the rotational temperatures, it is possible to estimate the kinetic
temperature ($T_{\rm kin}$). Since the derived $T_{\rm rot}$ values (see Table
5) are all around 10 K, the differences between kinetic and rotational
temperatures, as shown by Walmsley \& Ungerechts (\cite{walm}) and Danby et
al. (\cite{danby}), are within the errors of our $T_{\rm rot}$ measurements.
We thus can assume $T_{\rm kin}$ = $T_{\rm rot}$.

The excitation temperature ($T_{\rm ex}$), derived assuming a beam-filling
factor of one, is defined in an analogous way to $T_{\rm rot}$, but it is
related to the population across an inversion doublet (J,K). Using $T_{\rm
ex}$ and $T_{\rm rot}$, it is possible to derive the total ammonia column
density ($N_{\rm tot}$; cm$^{-2}$) following the equations given by Ungerechts
et al. (\cite{unghe2}). For this, we have neglected non-metastable (J $>$ K)
inversion doublets and have assumed a common rotational temperature for all
the metastable (J = K) states. These are reasonable approximations for
rotational temperatures around 10 K (e.g. Harju et al. \cite{harju}). The
derived temperatures and column density values (see Table 5) are consistent
with \amm observations towards other molecular cores in the TMC (e.g. Myers \&
Benson \cite{mybe}; Benson \& Myers \cite{bemy2}) and more isolated Bok
globules (Lemme et al. \cite{lemme}; Scappini \& Codella \cite{scappini}). The
measured (1,1) excitation temperatures are only slightly smaller than the
rotational temperatures (see Table 5) suggesting that the (1,1) transition is
close to being thermalised. $T_{\rm ex}$ $\simeq$ $T_{\rm rot}$ also requires
that the beam-filling factor is near unity. This implies that, for the 
sources of the present sample, most of the ammonia flux is arising from an
extended structure.

By balancing collisions and stimulated emission against spontaneous emission,
the hydrogen density ($n_{\rm H_{\rm 2}}$; cm$^{-3}$) can be estimated (Ho \&
Townes \cite{ho}). The derived $n_{\rm H_{\rm 2}}$ values, also accounting for
photon trapping, range from 0.6 10$^4$ to 19.9 10$^4$ cm$^{-3}$ (see Table 6)
and confirm that ammonia traces regions with densities $\ge$ 10$^4$ cm$^{-3}$.
Following Ho \& Townes (\cite{ho}), the errors on $n_{\rm H_{\rm 2}}$ are
17\% and 64\% for L1521D and L1507A, respectively. It is worth noting that the
errors associated with $n_{\rm H_{\rm 2}}$ become large when $T_{\rm ex}$ is
close to $T_{\rm rot}$. Thus, the hydrogen densities derived for L1521F and
L1524, i.e. for those sources with $T_{\rm ex}$ $\simeq$ $T_{\rm rot}$, may be
lower limits only.

It is well known that intrinsic motions in molecular clouds and, in
particular, in molecular cores are complex. The spectral lines of thermally
excited transitions also contain a non-thermal contribution to the linewidth
(e.g. Myers et al. \cite{mylafu}), which can provide information about the
dynamics of the molecular gas. In order to derive $\Delta v$, the {\it actual}
intrinsic linewidth of the (1,1) line, we have corrected the observed
intrinsic linewidth $\Delta v_{\rm INT}$ for the spectral resolution of the
autocorrelator used during the observations, $\Delta v_{\rm aut}$, by means of
$\Delta v^2 = \Delta v_{\rm INT}^2 - \Delta v_{\rm aut}^2$, assuming the
filter (and the line) being gaussian shaped. The actual intrinsic linewidth
$\Delta v$ is composed of a thermal component ($\Delta v_{\rm T}$) and a
non-thermal one ($\Delta v_{\rm NT}$): $\Delta v^2 = \Delta v_{\rm T}^2 +
\Delta v_{\rm NT}^2$ (e.g. Myers et al. \cite{mylafu}). It is possible to see
(Tables 2 and 5) that: (i) the actual intrinsic linewidths $\Delta v$ are
between 0.17 and 0.27 km s$^{-1}$ ($\Delta v_{\rm INT}$ $\simeq$ 0.28 km
s$^{-1}$) and (ii) the non-thermal velocity component is small, with a ratio
between $\Delta v$  and the thermal linewidth component $\Delta v$/$\Delta
v_{\rm T}$ $\simeq$ 1.3. The results of the \amm observations, performed by
Benson \& Myers (\cite{bemy2}) using the larger beamwidths of the Haystack and
Green Bank antennas, indicate that, on average, ammonia cores with embedded
IRAS source(s) have larger linewidths (0.45 km s$^{-1}$, not corrected for the
spectral resolution) than cores without IRAS point source (0.27 km s$^{-1}$).
Moreover, Myers et al. (\cite{mylafu}) studied the linewidths of 61 dense
condensations associated with star forming regions. They showed (see their
Fig. 1) that the non-thermal component of the \amm core motions increases with
IRAS luminosity more rapidly than does the thermal component and that in cores
which can form stars more massive than 2 $M_{\rm \sun}$, non-thermal motions
dominate thermal motions ($\Delta v$/$\Delta v_{\rm T}$ $\simeq$ 4). Thus, the
linewidths derived suggest that our \amm clumps are either not associated with
YSOs or that they are sites of low-mass star formation and are presumably
associated with IRAS point sources of low luminosity ($L_{\rm FIR}$ $<$ 10
$L_{\rm \sun}$; Myers et al. \cite{mylafu}), corresponding to a stellar mass
of less than 2 $M_{\rm \sun}$.

The LSR velocity information of the four mapped cores has been analysed in
order to search for velocity gradients: $v_{\rm LSR}$ does not vary greatly
within L1524 and L1507A, since the maximum differences between the $v_{\rm
LSR}$ values of the map positions are 0.16 km s$^{-1}$ and 0.14 km s$^{-1}$,
respectively. These values are comparable to the channel spacing of 0.154 km
s$^{-1}$. The map of L1521F shows a significant variation (0.46 km s$^{-1}$),
but there is no clear indication for a systematic velocity gradient along a
preferred axis. In contrast, L1521D displays a large variation of $v_{\rm
LSR}$ across the map revealing a velocity gradient roughly along the
north-south axis with a change of about 0.43 km s$^{-1}$ over 120$\arcsec$ 
(cf. Fig. 1; weak line emission from outside the outer 1.4\,K contour 
supports this trend). This corresponds to $\simeq$ 5 km s$^{-1}$ pc$^{-1}$,
which is a value definitely larger than those reported by Goodman et al.
(\cite{goo}) for dense cores ($<$ 4 km s$^{-1}$ pc$^{-1}$). It is also worth
noting that the direction of the gradient is not related to the major-axis of
the elongated \amm distribution of L1521D (Fig. 1). This is consistent with
the work of Goodman et al. (\cite{goo}), who studied the occurrence of
velocity gradients in a sample of 43 ammonia cores, suggesting that the motion
connected with the gradient, either rotation or shear, is not the major factor
in the core dynamics.

\begin{table*}
\caption[] {Derived parameters from \amm observations}
\begin{tabular}{l|r|r|r|r|r|r|r}
\hline 
\multicolumn{1}{c|}{{Name}} &
\multicolumn{1}{c|}{\bf {$T_{\rm rot}$}} &
\multicolumn{1}{c|}{\bf {$T_{\rm ex}$}} &
\multicolumn{1}{c|}{\bf {$N_{\rm 11}$}} &
\multicolumn{1}{c|}{\bf {$N_{\rm tot}$}} &
\multicolumn{1}{c|}{\bf {$\Delta v$}} &
\multicolumn{1}{c|}{\bf {$\Delta v_{\rm T}$}} &
\multicolumn{1}{c}{\bf {$\Delta v_{\rm NT}$}} \\
\multicolumn{1}{c|}{} &
\multicolumn{1}{c|}{(K)} &
\multicolumn{1}{c|}{(K)} &
\multicolumn{1}{c|}{(10$^{14}$~cm$^{-2}$)} &
\multicolumn{1}{c|}{(10$^{14}$~cm$^{-2}$)} &
\multicolumn{1}{c|}{(km~s$^{-1}$)} &
\multicolumn{1}{c|}{(km~s$^{-1}$)} &
\multicolumn{1}{c}{(km~s$^{-1}$)} \\ 
\hline
L1521D & 9.9 (1.2) & 6.0 (0.2) & 1.2 (0.3) & 5.4  (1.8) & 0.27 (0.02) & 0.16 (0.02) & 0.22 (0.02) \\ 
L1521F & 9.1 (1.0) & 9.0 (0.3) & 3.8 (0.3) & 20.1 (1.6) & 0.17 (0.01) & 0.16 (0.01) & 0.06 (0.01) \\
L1524  & 7.9 (0.7) & 7.5 (0.2) & 3.8 (0.3) & 27.7 (6.6) & 0.18 (0.01) & 0.15 (0.01) & 0.10 (0.01) \\
L1507A & 8.9 (1.7) & 6.8 (0.3) & 1.5 (0.2) & 8.3 (3.6)  & 0.20 (0.01) & 0.15 (0.02) & 0.13 (0.02) \\
\hline
\end{tabular}
\end{table*}

\subsection{IRAS counterparts}

It is well known that the IRAS PSC offers a precious opportunity to identify
star forming regions in optically obscured objects, such as molecular cores.
Using the IRAS PSC, Kenyon et al (\cite{kenyon}) have recently performed a
survey of star forming regions in the TMC, identifying about 100 YSOs. They
have shown that YSOs are scattered in a large area of 15 pc $\times$ 20 pc and
that star formation in the TMC is restricted to the dense molecular cores.
With the aim to collect more information about the nature of the four
molecular cores of the present list, we have cross-correlated their positions
with the IRAS PSC. Clark (\cite{clark}) has analysed the spatial distribution
of 396 IRAS PS located up to 2 pc from 60 molecular cores. He found that the
distribution is strongly peaked on the ammonia cores, with the central peak
reduced by an order of magnitude at 0.18 pc. At distances larger than this
value, a long tail is present and the population of IRAS PS can be
significantly affected by background sources. Considering that the typical
ammonia core diameter is 0.1 pc, 0.18 pc corresponds to 3.6 radii. Taking into
account this result, we have looked for a coincidence within a radius of 1.8
times the size of the four molecular cores of the present sample. Moreover,
IRAS sources known to be associated with extragalactic sources have been
excluded (IRAS \cite{iras}). The result of the cross-correlation shows that
L1521F and L1507A have no IRAS counterparts within the selected radius. We
find two IRAS PS around L1521D, IRAS\,04181+2655 (at a distance of 108$\arcsec$)
and IRAS\,04181+2654 (138$\arcsec$), and one near L1524, IRAS\,04263+2426
(120$\arcsec$). It is worth noting that all these IRAS PS satisfy the typical
IRAS colour distribution for IRAS counterparts of molecular cores, since they
have [25-12] ([i-j] $\equiv$ log [$F_{\rm i}$/$F_{\rm j}$], where $F_{\rm
i,j}$ are the IRAS flux densities in bands i and j, in $\mu$m) in the range
between 0.42 and 0.47 and [60-12] between 0.60 and 1.23 (for comparison, see
figure 3 of Codella \& Palla \cite{copa}).

We thus conclude that L1521D and L1524 are associated with at least one IRAS
counterpart (and therefore presumably with YSOs), while L1521F and L1507A
appear not to be connected with any IRAS PS. Following Henning et al.
(\cite{henning}), the IRAS luminosities of the associated IRAS PS has been
computed. The values of $L_{\rm FIR}$ are 0.3 $L_{\rm \sun}$ for
IRAS\,04181+2655 and IRAS\,04181+2654 (L1521D) and 6.1 $L_{\rm \sun}$ for
IRAS\,04263+2426 (L1524). This confirms that the ammonia molecular cores are
sites of low-mass star formation, in accord with our results from the ammonia
linewidths (Sect. 5.1).

\subsection{\amm core stability}

In order to obtain estimates of the energy terms of the \amm cores of our
sample, the {\it mean} \amm total column density has been obtained as the
average of all observed positions inside the FWHP contour of the ammonia cores
(Figs. 1--4). This allows to calculate, using the ammonia abundance and
assuming spherical cloud geometry, estimates of the {\it mean} hydrogen
density and of the mass $M$ ($M_{\rm \sun}$) of a molecular core (Table 6). It
is worth noting that, since $M$ has been derived from the \amm total column
density without virial equilibrium, we have made no assumption about the
stability of the \amm cores. Ammonia abundances in interstellar molecular
clouds cover a wide range: the [NH$_3$]/[H$_2$] ratio is believed to vary from
a few 10$^{-8}$ in small dark clouds up to 10$^{-5}$ in the dense cores like
Orion-KL (e.g. Ho \& Townes \cite{ho} and references therein). In particular,
Benson \& Myers (\cite{bemy0}) studied the \amm abundance in a sample of
quiescent dense molecular clouds and found that [NH$_3$]/[H$_2$] ranges
between 3 10$^{-8}$ and 2 10$^{-7}$.

In order to derive the parameters reported in this paper, we have assumed
[NH$_3$]/[H$_2$] = 10$^{-7}$. The resulting masses lie between 0.2 and 1.0
$M_{\rm \sun}$, placing the \amm cores with the lowest $M$ values (L1507A and
L1521D) towards the edge of the mass distribution for ammonia cores in the
TMC, which ranges from a fraction to a few tens of solar masses, with a median
value of 4 $M_{\rm \sun}$ (Benson \& Myers \cite{bemy2}).

Studying C$_3$H$_2$ and H$^{13}$CO$^+$ in TMC molecular cores, Mizuno et al.
(\cite{mizuno2}) found that, contrary to the \amm results of Benson \& Myers
(\cite{bemy2}), the cores without young stars (i.e. IRAS counterparts) are
less dense than those with stars. The results of Mizuno et al.
(\cite{mizuno2}) and Benson \& Myers (\cite{bemy2}) are consistent (average
values) regarding the hydrogen density of the cores not connected with stars
(4~10$^{4}$ and 3~10$^{4}$ cm$^{-3}$, respectively), while for those which
host YSOs Mizuno et al. (\cite{mizuno2}) find $n_{\rm H_{\rm 2}}$ $\simeq$
5~10$^{5}$ cm$^{-3}$, a value larger than that given by Benson \& Myers
(\cite{bemy2}), 1~10$^{4}$ cm$^{-3}$. It is worth noting that Mizuno et al.
(\cite{mizuno2}) determined the hydrogen density from H$^{13}$CO$^+$ data,
while the Benson \& Myers (\cite{bemy2}) analysis is based on a large velocity
gradient model for the \amm cores. Mizuno et al. (\cite{mizuno2}) suggest that
the disagreement of their results with those of Benson \& Myers (\cite{bemy2})
could be caused by differences in angular resolutions: 20$\arcsec$ vs.
1$\arcmin$.2, respectively. We stress that our results have
been obtained with a resolution of 40$\arcsec$, which is still lower that that
of the Mizuno et al. (\cite{mizuno2}) observations, but which is definitely
higher than that used by Benson \& Myers (\cite{bemy2}). The hydrogen densities
derived from our data show no remarkable difference between the $n_{\rm H_{\rm
2}}$ values of \amm cores with (L1521D and L1524) or without (L1521F and
L1507A) IRAS counterparts.

In Table 6 we display the energy terms of the cores of our sample: turbulent
($E_{\rm tu}$; J), thermal ($E_{\rm th}$; J) and gravitational ($E_{\rm gr}$;
J). The last column is for the parameter $\alpha$ $\equiv$ $E_{\rm
gr}$/(${E_{\rm tu} + E_{\rm th}}$), which is a measure of core stability.
The turbulent, thermal and gravitational (assuming a homogeneous sphere)
energy terms have been derived using the formulae reported by Harju et al.
(\cite{harju}).

\begin{table*}
\caption[] {Derived parameters related to the core stability of the \amm
condensations}
\begin{tabular}{l|r|r|r|r|r|r}
\hline 
\multicolumn{1}{c|}{{Name}} &
\multicolumn{1}{c|}{\bf {$n_{\rm H_{\rm 2}}$}} &
\multicolumn{1}{c|}{\bf {$M$}} &
\multicolumn{1}{c|}{\bf {$E_{\rm tu}$}} &
\multicolumn{1}{c|}{\bf {$E_{\rm th}$}} &
\multicolumn{1}{c|}{\bf {$E_{\rm gr}$}} &
\multicolumn{1}{c}{\bf {$\alpha$}} \\
\multicolumn{1}{c|}{} &
\multicolumn{1}{c|}{(10$^{4}$~cm$^{-3}$)} &
\multicolumn{1}{c|}{($M_{\rm \odot}$)} &
\multicolumn{1}{c|}{(10$^{34}$~J)} &
\multicolumn{1}{c|}{(10$^{34}$~J)} &
\multicolumn{1}{c|}{(10$^{34}$~J)} &
\multicolumn{1}{c}{} \\
\hline
L1521D &   0.6 & 0.3 &  0.9 &  3.5 &  0.9 & 0.2 \\ 
L1521F &  19.9 & 0.7 &  0.8 &  7.9 &  7.4 & 0.9 \\
L1524  &   2.5 & 1.0 &  1.0 &  8.9 & 11.8 & 1.2 \\
L1507A &   1.1 & 0.2 &  0.2 &  2.2 &  1.2 & 0.5 \\
\hline
\end{tabular}
\end{table*}

Star formation occurs most readily in molecular clumps in which the internal
motions are sufficiently reduced to allow the occurrence of gravitational
collapse. Table 6 shows that the thermal energy is always larger than the
turbulent one. The values of $\alpha$ for the four cores lie between 0.2
(L1521D) and 1.2 (L1524); $\alpha$ = 1 implies gravitational equilibrium in
the absence of a magnetic field. The values obtained here are only slightly
different from unity. Taking into account the contribution of a magnetic field
to the energetic support against gravitational collapse, considering also
uncertainties in the assumed [NH$_3$]/[H$_2$] value and deviations from the 
adopted (spherical) geometry of the cores, we can conclude that the four ammonia
cores are close to equilibrium. These considerations are in agreement with the
results of Myers \& Goodman (\cite{mygo}) and Harju et al. (\cite{harju}) who
found that the \amm cores in the TMC are approximately virialized.

\subsection{Derived parameters from HC$_5$N observations}

An estimate of the total HC$_5$N column density ($N_{\rm HC_5N}$; cm$^{-2}$)
has been derived through the standard equation (e.g. Olano et al.
\cite{olano}), using the molecular parameters given, e.g., by Alexander et al.
(\cite{alex}) and Churchwell et al. (\cite{chu}). Table 4 lists the derived
cyanodiacetylene column densities at the ammonia peak positions of the four
molecular cores: the values range from 1.6 10$^{12}$ to 9.2 10$^{12}$
cm$^{-2}$, in agreement with the $N_{\rm HC_5N}$ values of other molecular
cores located in the TMC (Benson \& Myers \cite{bemy0}).

In order to estimate the HC$_5$N abundances a comparison between HC$_5$N and
NH$_3$ column densities has been made. We presently do not know the size and
geometry of the HC$_5$N distribution in the ammonia cores. However, the
HC$_5$N and NH$_3$ maps given by Benson \& Myers (\cite{bemy0}) show that the
size and shape of the maps are similar and that the peak positions differ
significantly only for the TMC-1 molecular core. Therefore, we have 
estimated the [HC$_5$N]/[H$_2$] ratio using the cyanodiacetylene and ammonia
column densities and the NH$_3$ abundance. The obtained [HC$_5$N]/[NH$_3$]
values are: 8.3 10$^{-3}$ (L1521D), 4.6 10$^{-3}$ (L1521F), 2.7 10$^{-3}$
(L1524) and 10.7 10$^{-3}$ (L1507A). Thus, the derived HC$_5$N abundances 
relative to H$_2$ are: 8.3 10$^{-10}$ (L1521D), 4.6 10$^{-10}$ (L1521F), 2.7
10$^{-10}$ (L1524) and 1.1 10$^{-9}$ (L1507A), in agreement with the 
suggestion of Benson \& Myers (\cite{bemy0}), that relative cyanodiacetylene
abundances in dense cool cores vary from $\simeq$ 10$^{-10}$ to $\simeq$
10$^{-9}$.

\section{Conclusions}

We have discovered 4 dense cores (L1521D, L1521F, L1524 and L1507A) in the
TMC, increasing the number of known such objects to 20 in this prototypical
dark cloud complex. The main results of our study can be summarized as follows:

\begin{enumerate}

\item The ammonia observations reveal clumps with sizes of 0.06-0.09 pc and
yield kinetic temperatures between 7.9 and 9.9 K. The actual intrinsic
linewidths are between 0.17 and 0.27 km s$^{-1}$, while the thermal linewidths
are around 0.15 km s$^{-1}$. Therefore, the non-thermal velocity component is
rather small. The ratio between the intrinsic and thermal linewidth is about
1.3, indicating that only low-mass stars can eventually form in these \amm
cores. This is also supported by the association of two of the \amm cores with
low luminosity IRAS counterparts ($L_{\rm FIR}$ $<$ 10 $L_{\rm \sun}$).

\item The \amm data show that the measured (1,1) excitation temperatures are
only slightly smaller than the rotational temperatures suggesting that the
(1,1) inversion transition is close to being thermalised. From the peak \amm
emission hydrogen densities (between 0.6 and 19.9 10$^4$ cm$^{-3}$) have been
estimated. Moreover, from the average of all observed positions inside the
FWHP contour of the ammonia cores, the core masses (0.2 -- 1.0 $M_{\rm \sun}$)
have been estimated. The derived hydrogen densities show no remarkable
difference between \amm cores with or without IRAS counterparts. The low $M$
values place the four objects towards the edge of the $M$ distribution for
ammonia cores in the TMC, which ranges from fractions to few tens of solar
masses, with a median value of 4 $M_{\rm \sun}$.

\item Estimates of the turbulent, thermal and gravitational energy have been
derived. The thermal energy is always larger than
the turbulent one. The four cores may be close to equilibrium.

\item Emission at 24.0 GHz, caused by the J=9--8 transition of HC$_5$N, has been
detected towards the ammonia peak positions of the four molecular cores. This
increases the number of known such objects from 11 to 15 in the TMC (cf.
Benson \& Myers 1980, 1983). The derived HC$_5$N column densities range
between 1.6 10$^{12}$ and 9.2 10$^{12}$ cm$^{-2}$, in agreement with results
of previous HC$_5$N observations in the TMC.

\end{enumerate}

\vspace*{0.6cm}
\noindent
{\it Acknowledgements}
The authors like to thank M. Walmsley for valuable discussions and suggestions.

\end{document}